\documentclass[reprint, superscriptaddress, amsmath, amssymb, aps]{revtex4-2}

\usepackage{graphicx}
\usepackage{dcolumn}
\usepackage{bm}
\usepackage{array}
\usepackage{subcaption}
\usepackage{float}
\usepackage{textcomp}
\usepackage{bbding}
\usepackage{makecell}
\usepackage{siunitx}
\usepackage{upgreek}

\setcellgapes{4pt}

\begin{document}
	
	\preprint{APS/123-QED}
	
	\title{A novel apparatus for particle-particle single contact electrification experiments}
	
	\author{Otome Obukohwo}
	\affiliation{University of Ottawa, Canada}
	
	\author{Simon Jantač}
	\affiliation{Physikalisch-Technische Bundesanstalt, Germany.}
	
	\author{Andrew Sowinski}
	\affiliation{University of Ottawa, Canada}
	
	\author{Poupak Mehrani}
	\affiliation{University of Ottawa, Canada}
	
	\author{Holger Grosshans}
	\email{Corresponding author: Holger.Grosshans@ptb.de}
	\affiliation{Physikalisch-Technische Bundesanstalt, Germany.}
	\affiliation{Otto-von-Geuricke University Magdeburg, Germany.}
	
	\begin{abstract}
The experiment of a single contact between two sub-centimeter high-speed particles is often difficult to execute, especially if the collision must be physically and electrically isolated, as is the case for triboelectrification studies. Apparatuses designed for this type of experiment fall short of providing high-speed isolated collisions with a high probability of contact. In this article, we propose a novel apparatus that combines pneumatic conveying and acoustic levitation to provide an electrically and physically isolated, high impact speed collision between two sub-centimeter particles with a collision success rate of \qty{93}{\percent}. We can control the pre-contact charge, material, and size of both particles,and the impact speed and angle. Test results show that charge transfer between two insulator particles is not solely driven by contact potential difference; it is a stochastic process that requires large datasets to resolve and understand. Our new apparatus can efficiently generate these datasets and provide new insights on the stochastic nature of charge transfer, and the effect of each of the collision parameters mentioned earlier on particle-particle charge transfer.
\end{abstract}

	\maketitle
	
	\section{Introduction}
\label{section 01}

When two surfaces touch each other, there is a transfer of electrical charge between them. This phenomenon is known as triboelectrification, tribocharging, or contact charging. It is a major problem in industries involving transport, handling, or processing of small particles (or powders), as it causes unexpected particle flow patterns, cohesion among particles, adhesion to surrounding surfaces, and electrical discharges  - which could serve as ignition sources for dust explosions \cite{obukohwo_cfd_2023, supuk_tribo-electrification_2011, sowinski_comparison_2011, lacks_long-standing_2019}. Experimental studies aimed at understanding triboelectrification are divided into particle-plate (one curved and one flat surface) experiments and particle-particle (two curved surfaces) experiments. Particle-particle experiments are important for defining the mechanism of charge transfer between particles in industrial powders. The particles of interest are often very small (diameters ranging from a few micrometers to tens of millimeters \cite{ashrafi_particle_2008}) insulators that interact with each other at high speeds (gas velocity in pneumatic conveying up to \qty{30}{\meter\per\second} \cite{nimvari_velocity_2023}). Accordingly, the ideal apparatus for particle-particle experiments should allow for an electrically and physically isolated contact between two sub-centimeter particles at high impact speeds and, at a high collision success rate (high probability of contact between the particles).

An isolated contact is necessary to ensure that the contact point between the particles is the only point of charge transfer, and to ensure charge conservation; a high impact speed is necessary to recreate real-life industrial conditions; and a high collision success rate is necessary to ensure timely and efficient collection of data. The collision success rate is especially important because recently developed theories for charge transfer between insulator surfaces suggest that it is a stochastic process\cite{kok_electrification_2009, baytekin_material_2012}. Hence, large datasets are required to converge the statistical properties of any unique set of collision conditions. On that note, the ideal apparatus should also allow for variation of contact parameters that may affect triboelectrification. These parameters include, but are not limited to, particle size, contact surface material, impact speed, impact angle, and relative humidity and temperature of the surrounding gas  \cite{PhysRevMaterials.2.035602,saleh_relevant_2011,nimvari_effect_2022,kok_electrification_2009,baytekin_material_2012}.

\begin{table*}[tb]
\caption{\label{reported apparatuses} Reported apparatuses for particle-particle triboelectrification experimentation.}
\makegapedcells
\begin{ruledtabular}
\begin{tabular}{llccc}
\underline{}
    \makecell[l]{Authors} & 
    \makecell[l]{Collision and charge\\measurement techniques} & 
    \makecell[c]{High collision\\success rate?} & 
    \makecell[c]{Isolated\\contact?} & 
    \makecell[c]{High impact\\speed?} \\
    \hline
    
    \makecell[l]{Chowdhury\\\textit{et al.} \cite{chowdhury_apparatus_2020, chowdhury_electrostatic_2021}} & 
    \makecell[l]{Converging gas jets\\Faraday cage} &
    \XSolid &
    \Checkmark &
    \Checkmark  \\ 
    
    \makecell[l]{Han \\\textit{et al.} \cite{han_charging_2021}} &
    \makecell[l]{Physical supports for particles\\Faraday cage} &
    \Checkmark &
    \XSolid &
    \Checkmark \\

    \makecell[l]{Kline \\\textit{et al.} \cite{kline_precision_2020}} &
    \makecell[l]{Acoustic levitation\\Electric field} &
    \Checkmark &
    \Checkmark &
    \XSolid \\

    \makecell[l]{Jungmann \\\textit{et al.} \cite{jungmann_violation_2021}} &
    \makecell[l]{Micro gravity\\Electric field} &
    \makecell[c]{N/A} &
    \Checkmark &
    \XSolid  \\
    
\end{tabular}
\end{ruledtabular}
\end{table*}

There are notable apparatuses designed specifically for particle-particle experiments (Table~\ref{reported apparatuses}), but they fall short of the ideal apparatus. Chowdhury \textit{et al.} \cite{chowdhury_apparatus_2020,chowdhury_electrostatic_2021} used two converging gas jets to force two particles (\qtyrange{3.2}{4.8}{\milli\meter} spheres; polytetraflouroethylene (PTFE), aluminum, and nylon) toward each other for a collision. They dropped the particles through hollow Faraday cages, to measure their pre-contact charges, and into the converging gas jets which forced the particles to collide with each other. After the collision, the particles were caught in Faraday cups to measure their post-contact charges. A high-speed recording of each experiment was analyzed to confirm the collision. This set-up provided isolated contacts at high impact speeds, but with a very low collision success rate. The unpredictable motion of the gas jets made it more likely for particles to miss each other than collide with each other.

Han \textit{et al.} \cite{han_charging_2021} fixed one particle (incident sphere; \qty{5}{\milli\meter}; PTFE, polyformaldehyde (POM), or polyamide 66 (PA 66)) on an insulating string connected to a sliding rail. This incident sphere was dropped on another particle (target sphere; \qty{10}{\milli\meter}; PTFE, POM, or PA 66) which was fixed on an insulating stick inside a Faraday cage. The incident sphere fell into the Faraday cage, made contact with the target sphere, and left the cage. The Faraday cage measured the pre-contact charges of both particles and the post-contact charge of the incident sphere. This set-up provided high impact speeds and a high collision success rate, but the contacts were not isolated. The insulating stick on the target particle and the insulating string on the incident particle eliminated physical and electrical isolation.

Kline \textit{et al.} \cite{kline_precision_2020} fixed two particles (two \qty{710}{\micro\meter} - \qty{850}{\micro\meter} polyethylene spheres, two \qty{620}{\micro\meter} - \qty{780}{\micro\meter} polystyrene spheres, and one polystyrene sphere with a \qty{620}{\micro\meter} - \qty{780}{\micro\meter} sulfonated polystyrene sphere) in an acoustic wave trap and changed the location of the acoustic pressure nodes to force the particles toward each other for a collision and return them to their original positions after the collision. A controlled alternating electric field forced the particles to oscillate pre- and post-contact and a high-speed camera recorded the trajectories of the oscillating particles. Individual particle charges were estimated through analysis of the particle oscillation and its force balance. This set-up provided isolated contacts at a high collision success rate, but the impact speeds were low ($\approx$ \qty{0.05}{\meter\per\second} from reported figures).

Jungmann \textit{et al.} \cite{jungmann_violation_2021} injected \qty{0.1}{\milli\gram} of \qty{434}{\micro\meter} glass grains into a micro gravity chamber ($<10^{-5}$g) and recorded particle collisions with a high-speed camera. Each particle's motion pre- and post-contact was influenced by a controlled electric field in the micro-gravity chamber. Individual particle charges were estimated through analysis of the particle's motion and its force balance. This set-up provided isolated contacts but with low impact speeds ($\approx$ \qty{0.2}{\meter\per\second} from reported figures) and an indeterminable collision success rate - it is safe to assume that there were some contacts involving more than two grains or the walls of the micro gravity chamber.

There is still no apparatus that simulates an electrically and physically isolated contact between two particles at high impact speeds and, at a high collision success rate. To fill this gap, we propose an apparatus that combines pneumatic conveying and acoustic levitation to provide high speed isolated collisions with a high collision success rate. Faraday cages measure the pre- and post-contact particle charges and high-speed cameras record the particle trajectories to confirm collision and provide estimates for impact angle and speed. In a successful experiment, the pneumatic conveyor propels an incident particle, with controlled speed, trajectory, and pre-contact charge, towards a target particle held stationary in an acoustic trap. After the collision, either both particles move forward into the post-contact measurement section, or the target particle moves into the post-contact charge measurement section while the incident particle stays in the acoustic trap. In the latter case, the trap is eliminated and the incident particle falls through an opening at the center of the reflector and into a Faraday cup where its post-contact charge is measured.

	\section{Apparatus design and control of physical collision conditions}
\label{section 02}

Figure~\ref{apparatus schematic} shows a schematic of our apparatus. The apparatus is divided into three sections: the pre-contact section, the contact section, and the post-contact section. 

\begin{figure*}[tb]
	\centering
	\includegraphics[width = \textwidth]{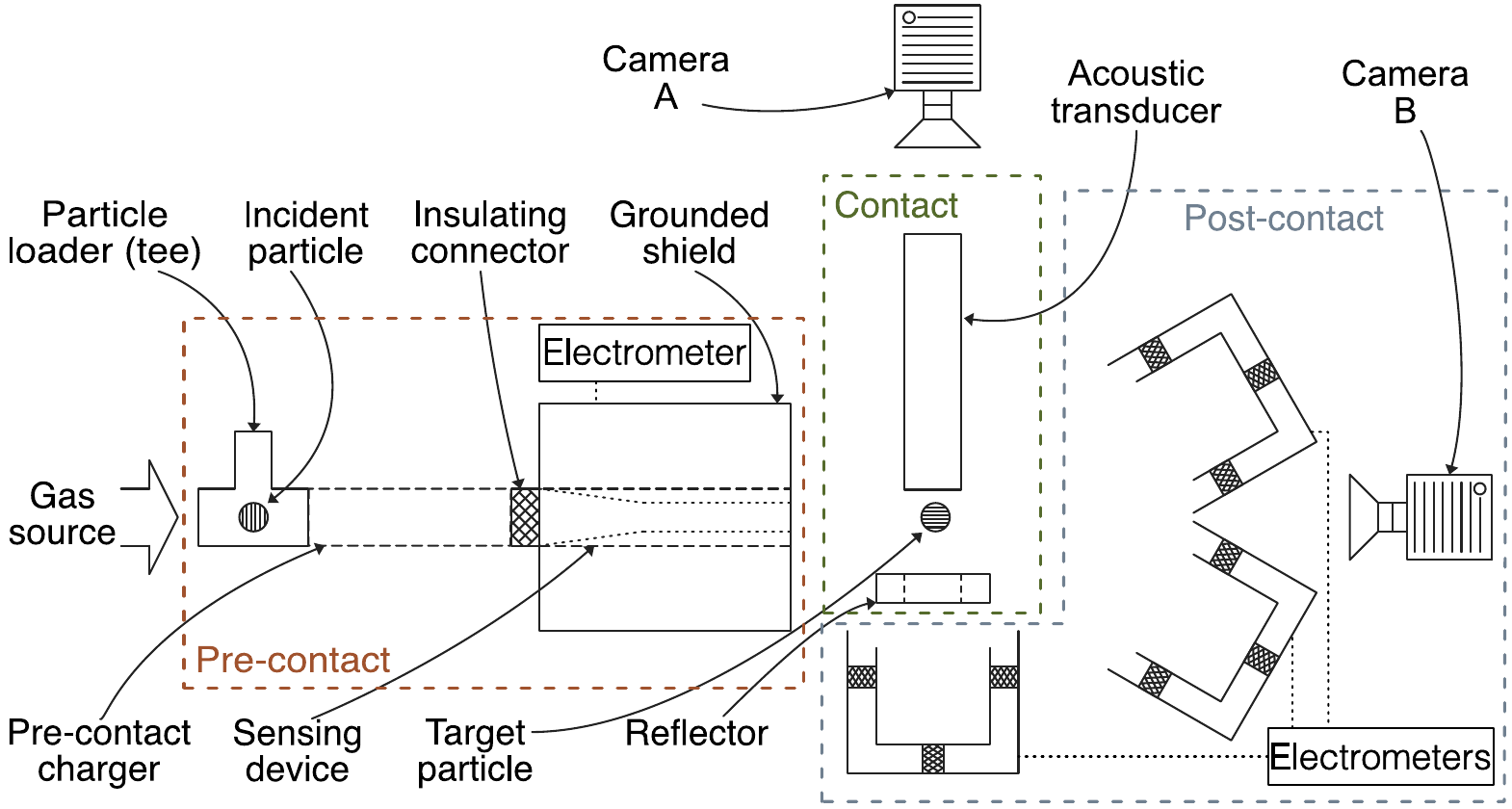}
	\caption{Schematic of our apparatus. The pre-contact section propels an incident particle, with controlled speed, trajectory, and pre-contact charge, towards a target particle held stationary in the contact section. After the collision, the post-contact charge of both particles is measured by Faraday cups in the post-contact section. This apparatus combines pneumatic conveying and acoustic levitation to provide an electrically and physically isolated, high impact speed collision between two sub-centimeter particle with a collision success rate of \qty{93}{\percent}. We can precisely control the incident pre-contact charge, incident trajectory, impact speed, impact angle, and target pre-contact charge.}
	\label{apparatus  schematic}
\end{figure*}

The pre-contact section controls the pre-contact charge, trajectory and speed of the incident particle. This section consists of a particle loader, a pre-contact charger, and a sensing device. The particle loader, a stainless-steel 316 1/4-inch union tee with a cap on the third connection port, is the primary position of the incident particle. At the onset of gas flow into the particle loader, the incident particle proceeds to the pre-contact charger, where it is charged before contact. The pre-contact charger consists of one or two coils (inner diameter: \qty{55}{\milli\meter}; free length: \qty{330}{\milli\meter}; revolution count: 15.5), each made from a 1/4-inch stainless-steel 316 tube. The charger can be eliminated to minimize pre-contact charging of the incident particle. After charging, or not, the incident particle moves into the sensing device where its pre-contact charge is measured, and it is given a controlled trajectory toward the target particle. The sensing device consists of three concentric tubes: a \qty{35}{\milli\meter} copper tube (shield), a 1/2-inch stainless-steel 316 tube (sensor), and a \qty{10}{\milli\meter} plastic tube (reducer). The gas flow rate controls the incident impact speed, the pre-contact charger controls the incident pre-contact charge, and the sensing device controls the incident particle trajectory.

\begin{figure*}[tb]
\centering
\begin{subcaptionbox}{\label{incident trajectory}}[0.49\textwidth]
    {\includegraphics[width=\linewidth]{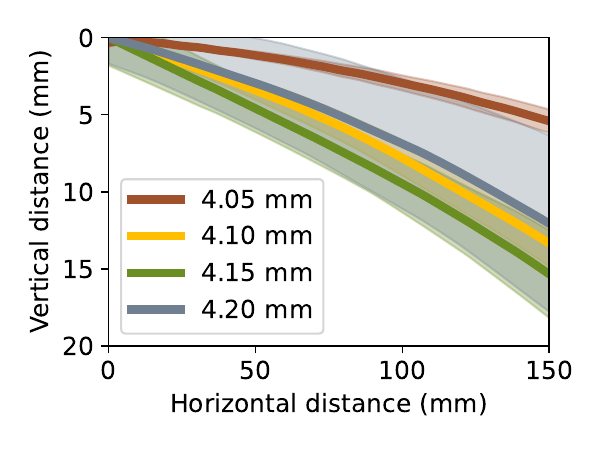}}
\end{subcaptionbox}
\begin{subcaptionbox}{\label{planar precision}}[0.49\textwidth]
    {\includegraphics[width=\linewidth]{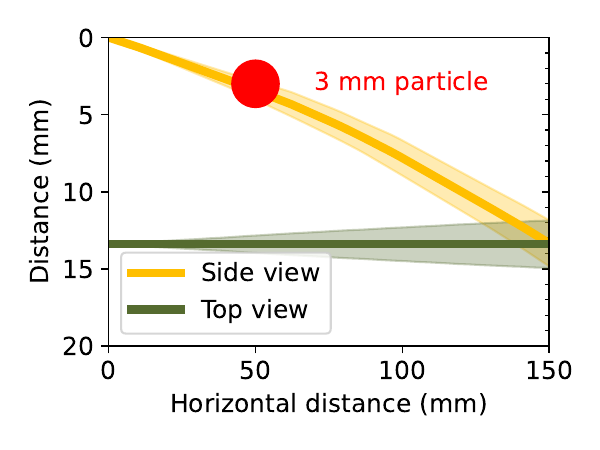}}
\end{subcaptionbox}
\caption{Mean incident particle trajectory (with one standard deviation as shaded region) for \qty{3.97}{\milli\meter} nylon sphere through reducers with different outlet diameters shown in (A). The precision of the trajectory increases as $\lambda$ decreases, and the precision is the same in all planes (B). The red dot in (B) represents a \qty{3}{mm} target particle \qty{50}{mm} away from the reducer outlet. The spread of the incident trajectory is less than the particle diameter, so the incident particle has a near-perfect chance of hitting this target particle.  This result confirms that we have precise control the incident particle trajectory towards the target particle.}
\label{}
\end{figure*}

For precise and repeatable incident particle trajectories, the reducer in the sensing device has an outlet inner diameter ($ID_{outlet}$) that is less than its inlet inner diameter (\qty{8}{\milli\meter}) and is dependent on the incident particle diameter ($D_{p,i}$) as follows:
\begin{equation}
ID_{outlet} = \left(\lambda+1\right)D_{p,i}.
\label{reducer outlet diameter}
\end{equation}
To decide the optimal $\lambda$ for each particle diameter, we recorded the trajectory of a \qty{3.97}{\milli\meter} nylon sphere conveyed with a gas velocity of \qty{20}{\meter\per\second} through reducers with different outlet diameters ($\lambda =$ 0.02, 0.03, 0.05, and 0.06). The mean trajectory for each $\lambda$ plotted along with one standard deviation (shaded region around the mean line) is shown in Figure~\ref{incident trajectory}. The deviation of the particle reduces with $\lambda$ and at $\lambda =$ 0.03, the trajectory deviation 150 mm away from the reducer outlet is less than one particle diameter. To check if the precision of the incident particle trajectory is the same in all planes, we recorded the trajectory of the \qty{3.97}{\milli\meter} nylon sphere through the $\lambda =$ 0.03 reducer in two perpendicular planes and plotted the mean and deviation of the trajectories in Figure~\ref{planar precision}. The perfect overlap of the deviations at a horizontal distance of \qty{150}{\milli\meter} suggests the precision of the incident particle trajectory is identical in all planes. The red dot in Figure~\ref{planar precision} represents a \qty{3}{mm} target particle \qty{50}{mm} away from the reducer outlet. The spread of the incident trajectory is less than the particle diameter, so the incident particle has a near-perfect chance of hitting this target particle.  This result confirms that we have precise control the incident particle trajectory towards the target particle.

\begin{figure}[tb]
\centering
\includegraphics[width=0.45\textwidth]{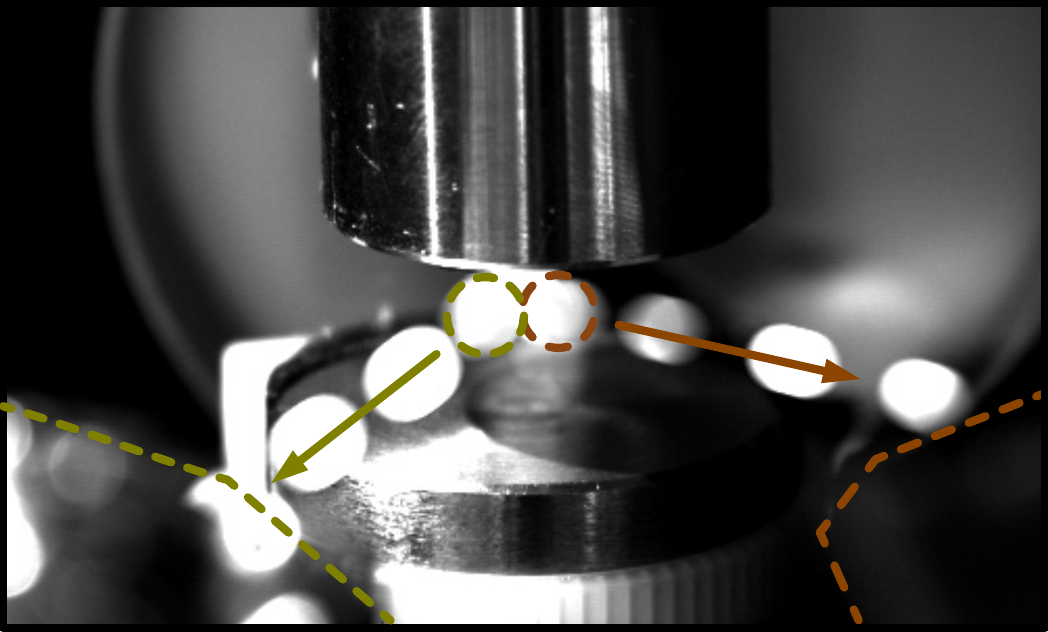}
\caption{Overlapped frames of a collision of a \qty{3.97}{\milli\meter} nylon sphere (incident particle in green) and a \qty{3.17}{\milli\meter} nylon sphere (target particle in brown). The borders of the Faraday cups are highlighted consistently with the colour of the particle the cup catches. The precision of the incident particle trajectory allows us to control the impact angle and post-contact trajectories.}
\label{collision frames}
\end{figure}

The contact section controls the pre-contact charge and position of the target particle. This section consists an ultrasonic transducer and a metal reflector. The ultrasonic transducer is a Hielscher\textsuperscript{\textregistered} UIP 500hdt with BS4d22 and BS4-2.2 adapter, and the metal reflector is a stainless-steel cylinder (dimater $=$ \qty{12}{mm}). The distance between the transducer and the reflector is set to a multiple of half the generated sound wavelength to create an acoustic trap that holds the target particle stably in place for a collision. Figure~\ref{collision frames} shows overlapped frames of one collision between a \qty{3.97}{\milli\meter} nylon incident sphere and a \qty{3.17}{\milli\meter} nylon target sphere. The particle, its trajectory, and the border of the Faraday cup opening are highlighted in brown for the target particle and green for the incident particle. The reflector diamter is small that it does not obstruct the path of the particles moving into the post-contact section. Therefore, we can hold a physically and electrically isolated target particle in place, and both incident and target particles do not touch anything after the collision and before they enter the post-contact section ensuring charge conservation between both particles during an experiment.

After a collision, each particle moves into the post-contact section where its post-contact charge is measured. If both particles move forward after the collision, two Faraday cups (one for each particle) collect the particles and measure their charges. If the target particle moves forward while the incident particle stays in the acoustic trap, the trap is eliminated and the incident particle falls through a hole in the reflector into a Faraday cup that collects it and measures its charge.

The combination of the precise repeatable incident particle trajectories and the stable levitated particle provides collisions with high impact speeds, physical and electrical isolation, and at high collision success rate of \qty{93}{\percent}. Our apparatus meets all the criteria for experimentation of particle-particle single contact charge transfer; this is a significant improvement from the apparatuses reported in literature.

	\section{Electrical collision conditions and impact charge tests}
\label{test results}

\begin{figure*}[tb]
	\centering
	\begin{subcaptionbox}{\label{incident pre-contact charge}}[0.49\textwidth]
		{\includegraphics[width=\linewidth]{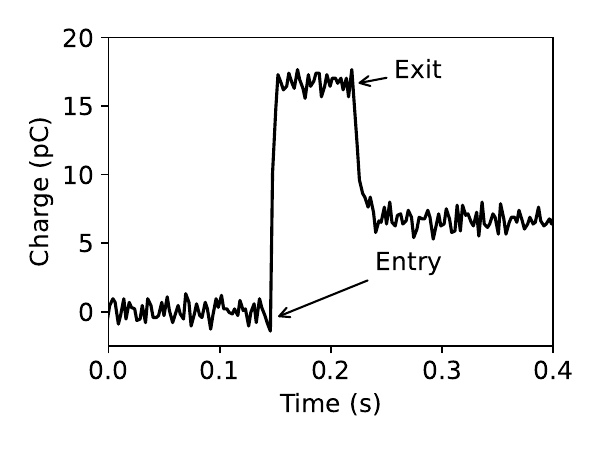}}
	\end{subcaptionbox}
	\begin{subcaptionbox}{\label{sensing device charge comparison}}[0.49\textwidth]
		{\includegraphics[width=\linewidth]{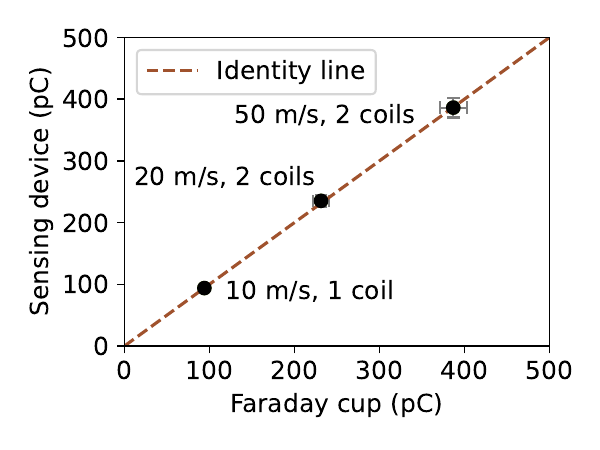}}
	\end{subcaptionbox}
	\caption{(A) Incident pre-contact charge measurement via the sensing device for a neutralized \qty{3.97}{\milli\meter} nylon sphere conveyed with zero coils in the pre-contact charger. The incident pre-contact charge is equal to the difference between the mean charge value after exit of the incident particle and the mean charge value during residence of the incident particle in the sensor. (B) Comparison between incident pre-contact charge measurements from the sensing device and a Faraday cup for the same \qty{3.97}{\milli\meter} nylon sphere conveyed with a one or two coils at different gas velocities. All data points fall on the identity line, confirming that the sensing device precisely and accurately measures the incident pre-contact charge.}
	\label{}
\end{figure*}

\begin{figure*}[tb]
	\centering
	\begin{subcaptionbox}{\label{neutralization test}}[0.49\textwidth]
		{\includegraphics[width=\linewidth]{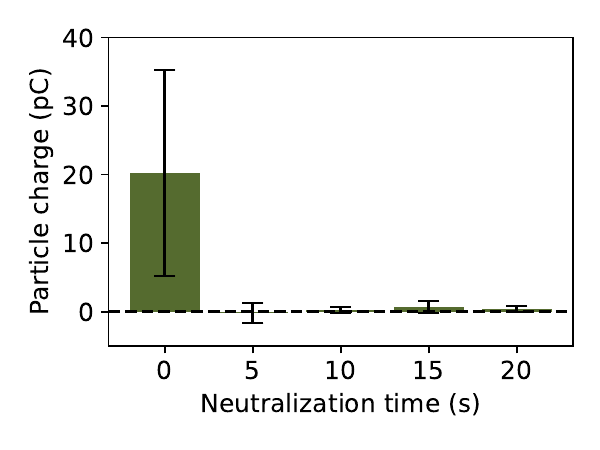}}
	\end{subcaptionbox}
	\begin{subcaptionbox}{\label{charging channel}}[0.49\textwidth]
		{\includegraphics[width=\linewidth]{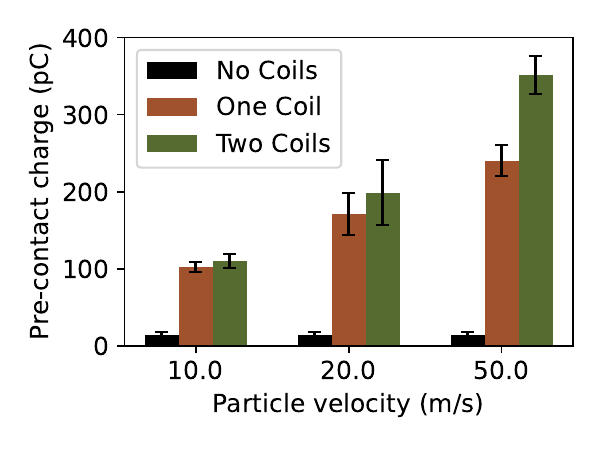}}
	\end{subcaptionbox}
	\caption{(A) Particle charge of a \qty{3.17}{\milli\meter} nylon sphere after neutralization for different durations. The magnitude and error of the post-neutralization charges (\qty{0.4}{\pico\coulomb} $\pm$ \qty{0.4}{\pico\coulomb}) confirm that we can precisely control the target pre-contact charge. (B) Incident pre-contact charge as a function of particle velocity and number of coils in the pre-contact charger. The data shows that we can precisely control the incident pre-contact charge with a maximum deviation of \qty{20}{\percent}.}
	\label{preliminary results}
\end{figure*}

For accurate and precise incident pre-contact charge measurement, the shield in the sensing device is grounded, the sensor is completely enclosed in the shield, and the reducer is made out of plastic - so it does not interfere with the field lines of the incident particle. To test the precision and accuracy of the sensing device, we conveyed neutralized \qty{3.97}{\milli\meter} spheres (neutralization tests are discussed later in this section) through the pre-contact charger (with zero, one, or two coils), through the sensing device, and into a Faraday cup at \qty{10}{\meter\per\second}, \qty{20}{\meter\per\second}, and \qty{50}{\meter\per\second}. Figure~\ref{incident pre-contact charge} shows a sample of the data from the sensing device connected to a Keithley\textsuperscript{\textregistered} 6514 electrometer for a case with zero coils. The incident pre-contact charge is equal to the difference between the mean charge value after exit of the incident particle and the mean charge value during residence of the incident particle in the sensor. Figure~\ref{sensing device charge comparison} shows a comparison between the incident pre-contact charge measurements from the sensing device and the Faraday cup. All data points fall on the identity line, confirming that the sensing device precisely and accurately measures the incident pre-contact charge.

\begin{figure*}[tb]
	\centering
	\begin{subcaptionbox}{\label{incident impact}}[0.49\textwidth]
		{\includegraphics[width=\linewidth]{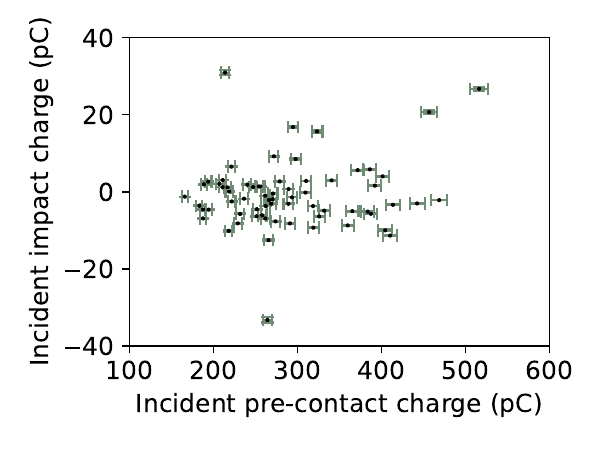}}
	\end{subcaptionbox}
	\begin{subcaptionbox}{\label{target impact}}[0.49\textwidth]
		{\includegraphics[width=\linewidth]{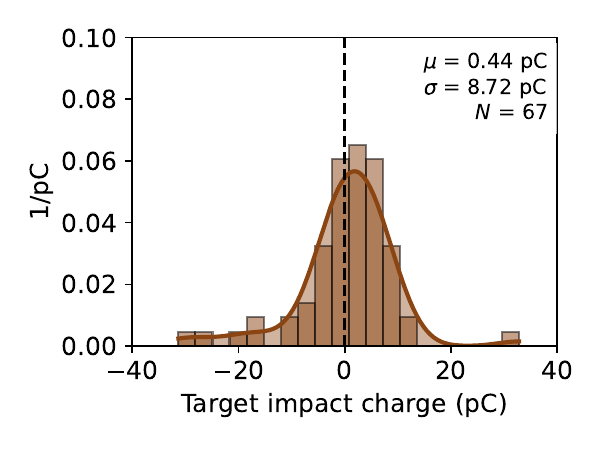}}
	\end{subcaptionbox}
	\caption{(A) Probability density distribution of the incident impact charges. The data shows that our high collision success rate allows us to obtain enough data points to resolve charge transfer statistics. Analysis of the incident impact charge as a function of the incident initial charge (B) suggests that all collision variables should be controlled and kept constant until converged statistics are obtained for each unique set of collision conditions.}
	\label{preliminary results}
\end{figure*}

To test our control of the target pre-contact charge, we levitated \qty{3.17}{\milli\meter} nylon spheres, neutralized them for different durations with a SIMCO\textsuperscript{\textregistered}-ION MPM UL 25 R ion-generator, and dropped them through the hole in the reflector into a Faraday cup to measure their charge. Figure~\ref{neutralization test} shows the results of the neutralization test. After neutralizing for 20 seconds, the particle charge reduced from \qty{20.3}{\pico\coulomb} $\pm$ \qty{15.0}{\pico\coulomb} to \qty{0.4}{\pico\coulomb} $\pm$ \qty{0.4}{\pico\coulomb}. These tests confirm that we can precisely control the target pre-contact charge.

To test our control of the incident pre-contact charge, we conveyed a neutralized (for 20 seconds) \qty{3.97}{\milli\meter} nylon sphere through zero, one, and two coils, at  \qty{10}{\meter\per\second}, \qty{20}{\meter\per\second}, and \qty{50}{\meter\per\second}. Figure~\ref{charging channel} shows the measured incident pre-contact charges. The pre-contact charge magnitude increased with the number of coils, and with the particle velocity. Addition of the first coil increases the particle charge five to eight times but addition of a second coil does not increase the particle charge significantly at speeds equal to and less than \qty{20}{\meter\per\second}. The effect of particle velocity is consistent with literature \cite{watanabe_new_2007,xu_experimental_2023,matsusaka_electrification_2000}. The observed effect of the number of coils is as expected - the number of coils is directly proportional to the number of particle-wall collisions which is, in turn, directly proportional to the magnitude of charge transfer between the incident particle and the wall. These tests confirms that we can precisely control the incident pre-contact charge with a maximum deviation of \qty{20}{\percent}.

To test the apparatus as a whole, we conveyed a \qty{3.97}{\milli\meter} nylon sphere (incident particle) at \qty{20}{\meter\per\second} towards a \qty{3.17}{\milli\meter} nylon sphere (target particle). The target particle was neutralized before every collision; the incident particle was not neutralized to induce a variation in the incident pre-contact charge. Figure~\ref{incident impact} shows the incident impact charge as a function of the incident initial charge. The data shows that the incident impact charge polarity could be positive or negative regardless of the magnitude and polarity of the incident pre-contact charge. This suggests that charge transfer between surfaces of identical materials is more than just a transfer of pre-contact charge driven by a contact potential difference, contrary to the suggestion of the condenser model \cite{matsusaka_electrostatics_2003}. The charge transfer may be dominated by physio-chemical processes at the surfaces in contact.  Figure~\ref{target impact} shows the distribution of the target impact charge as a probability density plot and a histogram. The positive mean of the distribution suggests that on average, the target particle gains charge from the incident particle., as expected. The gaussian distribution of impact charge is consistent with literature\cite{grosshans_unifying_2025}, and the wide spread of the distribution is further proof that the charge transfer between identical surface materials is a stochastic process, consistent with the surface state model \cite{kok_electrification_2009, baytekin_material_2012} and other experimental results \cite{PhysRevMaterials.3.085603,grosshans_unifying_2025,PhysRevLett.130.098202}. The stochastic nature of charge transfer explains the lack of any apparent trends in Figure~\ref{incident impact}: all collision variables should be controlled and kept constant until converged statistics are obtained for each unique set of collision variables - a feat that can be achieved with our apparatus.

	\section{Conclusions}
\label{conclusion}

Understanding a fundamental, and historically confusing, phenomenon like charge transfer requires a larger dataset, hence the motivation for our design and development of the novel apparatus we present in this paper. Our apparatus allows for an electrically and physically isolated, high impact speed collision between two sub-centimeter particle with a collision success rate of 93 \%. We are able to precisely control the incident pre-contact charge, incident trajectory, impact speed, impact angle, and target pre-contact charge. We show that large datasets are required to resolve the stochastic nature of charge transfer and study the effect of collision conditions. Our apparatus can be used to build these datasets and study the effects of pre-contact charge (as we have premiered in this paper), impact speed and angle, and relative humidity and temperature of the surrounding gas. These datasets will inform charge transfer estimation models and allow for better prediction and mitigation of the hazards of triboelectrification in industry.
	
	\section{acknowledgments}
\label{acknowledgments}

This project was funded by the European Research Council (ERC) and the National Sciences and Engineering Research Council of Canada (NSERC). ERC funding was provided by the European Union's Horizon 2020 research and innovation program (grant agreement number 947606 PowFEct).
	
	\bibliography{references}
	
\end{document}